\documentclass{article}
\usepackage{cite}
\newcommand{\pr}{{^\prime}}

\newcommand{\p}[1]{(\ref{#1})}

\newcommand{\cF}{{\cal F}}

\newcommand{\ba}{{\bar a}}
\newcommand{\bb}{{\bar b}}
\newcommand{\bc}{{\bar c}}
\newcommand{\bd}{{\bar d}}

\newcommand{\bm}{{\bar m}}
\newcommand{\bn}{{\bar n}}
\newcommand{\bp}{{\bar p}}

\newcommand{\bq}{{\bar q}}

\newcommand{\be}{\begin{equation}}
\newcommand{\ee}{\end{equation}}
\newcommand{\bea}{\begin{eqnarray}}
\newcommand{\eea}{\end{eqnarray}}

\newcommand{\nn}{\nonumber}

\usepackage{amscd,amsmath,amssymb}

\begin{document}
\thispagestyle{empty}
\vspace{2cm}
\begin{flushright}
\end{flushright}\vspace{2cm}
\begin{center}
{\Large\bf On interacting non-Abelian antisymmetric tensor field
models}
\end{center}
\vspace{1cm}

\begin{center}
{\large\bf
I.~Buchbinder${}^{a,b}$\footnote{buchbinder@theor.jinr.ru},
N.~Kozyrev${}^a$}\footnote{nkozyrev@theor.jinr.ru}
\end{center}

\begin{center}
${}^a$ {\it Bogoliubov  Laboratory of Theoretical Physics, JINR,
141980 Dubna, Russia}

${}^b$ {\it Center of Theoretical Physics, Tomsk State Pedagogical
University, 634061, Tomsk, Russia} \vspace{0.2cm}

\end{center}
\vspace{2cm}

\begin{abstract}\noindent
We study the gauge structure of interacting antisymmetric tensor field
models obtained by the dimensional reduction on a group manifold. In
such models a dimensionful parameter related to the size of the compact manifold plays
the role of a coupling constant. We focus on open issues of
reducibility of the gauge transformations in these theories and show that stages of reducibility of the
interacting models coincide with ones of corresponding free
counterparts.

\end{abstract}

\setcounter{page}{1}
\setcounter{equation}{0}

\section{Introduction}
Gauge fields are the natural elements of all approaches to
unification of fundamental interactions including gravity. It is
quite expected that development of new models of gauge fields may
present new possibilities for the theory of fundamental
interactions. In this regard, the study of bosonic and fermionic
antisymmetric tensor fields is of some interest.

Antisymmetric tensor fields or $p$-forms are the inherent
ingredients of the superstring/brane theory and supergravity models
in various dimensions (see, e.g.,\cite{GSchW,P,J,O,BBSch,FVP,BLT,Ta}). They are
characterized by a non-standard complicated gauge structure, which
may yield the problems for quantizing such field theories. The
simplest example of the antisymmetric tensors is the $2$-form field
which was introduced by Ogievetsky and Polubarinov \cite{OP}, who
showed that it is dual to free real scalar field
model,\footnote{Model of massive antisymmetric tensor field was
introduced earlier by Kummer \cite{KUMM} as the dual to the free
massive vector field model} and later was rediscovered in the string
theory context \cite{KR,CSc,GScO}. It is described
by the potential $B_{\mu\nu}$ with the gauge transformation law
 \be
  \label{Babelian}
\delta_{\Lambda} B_{\mu\nu} =
\partial_{\mu}\Lambda_{\nu}-\partial_{\nu}\Lambda_{\mu}.
 \ee
The field strength of $B_{\mu\nu}$, invariant with respect to
\p{Babelian}, is
 \be
  \label{strength}
  F_{\mu\nu\rho} = \partial_{\mu}
B_{\nu\rho}-\partial_{\nu} B_{\mu\rho}+\partial_{\rho} B_{\mu\nu},
 \ee
and the simplest (free) invariant action in $d$ dimensions is
 \be
 \label{2-formaction}
 S_{0}[B] = -\frac{1}{12}\int d^d x
F_{\mu\nu\rho}F^{\mu\nu\rho}.
 \ee
A peculiarity of \p{Babelian} gauge transformations is that they
can be trivial for a non-zero parameter, a property not found in electromagnetism, Yang-Mills
theory and gravity. In this case the corresponding parameter of a trivial transformation
is $\Lambda_\mu =\partial_\mu f$. Moreover, two transformations with different
parameters $\Lambda_{\mu}$ and $\Lambda{}^{(f)}_{\mu} =\Lambda_{\mu}+\partial_\mu f$ are identical, and the gauge parameter is
effectively defined up to its own gauge transformation. Such
transformations are called reducible \cite{BV}, and \p{Babelian} are
of the first stage of reducibility.

For higher $p$-forms, gauge structure becomes more complicated. The
transformation of three-form $C_{\mu\nu\rho}$
 \be
 \label{Babelian3}
 \delta_\Lambda C_{\mu\nu\rho} =
\partial_{\mu}\Lambda_{\nu\rho} - \partial_\nu \Lambda_{\mu\rho} +
\partial_{\rho}\Lambda_{\mu\nu}
 \ee
 with the parameter $\Lambda_{\mu\nu}$ coincides with the transformation with the parameter
 \be
  \label{Babelian3amb}
\Lambda{}^{(f)}_{\mu\nu}=\Lambda_{\mu\nu} +  \partial_\mu f_\nu -
\partial_\nu f_\mu.
 \ee
However, $\Lambda{}^{(f)}_{\mu\nu}$ are themselves ambiguous, as
 \be
 \label{2ambig}
 \Lambda{}^{(f^\prime)}_{\mu\nu} =
\Lambda{}^{(f)}_{\mu\nu} \;\; \mbox{if} \;\; f^{\prime}_\mu = f_\mu
+ \partial_\mu h
 \ee
 for some $h(x)$. Therefore, the gauge
transformations of the $3$-form are of second stage of reducibility.
For general $p$-form the stage is $p-1$ \cite{KUZ1}.

The gauge structure of corresponding field models poses a challenge
for quantization methods (see for review, e.g., the recent papers
\cite{KUZ1,KUZ2,BBKN1,BBKN2,ABBKN}
and references therein). Simple covariant quantization of Yang-Mills
type gauge theories is usually performed following the Faddeev-Popov
procedure \cite{FP}, which involves introduction of gauge fixing
conditions and ghost fields to construct the path integral of the
theory to ensure that integration effectively occurs not over all
the field configurations but over the orbits of the gauge group.
However, in the case of reducible theories this construction fails,
as the Faddeev-Popov-DeWitt determinant, which enters the path
integral as a factor, vanishes for reducible gauge transformations.
General covariant quantization method for reducible gauge theories
was developed by Batalin and Vilkovisky \cite{BV}. Recently there
was proposed a general enough simplified procedure for quantization
of reducible gauge theories based on generalization of Faddeev-Popov
procedure \cite{BBKN1,BBKN2,ABBKN} where the analogs
of Faddeev-Popov-DeWitt determinants were derived.\footnote{Such a
procedure was firstly used in work \cite{BK} for deriving the
effective actions for reducible ${\cal N}=1$ superfield theories in
background of superfield supergravity.}

The question of $p$-form field interactions are complicated on its
own, with one of the discussed possibilities being the non-Abelian
generalization. Though no-go theorems \cite{henneaux1},
\cite{bekaert1} imply that it is not possible to deform
transformation laws \p{Babelian} in non-Abelian way if the theory is
local and no extra fields are added, search for non-Abelian tensor
models continues, primarily motivated by the properties of stacks
$M5$-branes \cite{witten,witten2,witten3} and the
fact that the ${\cal N}=(2,0)$ superconformal field theory, the
maximal superconformal theory possible \cite{nahm}, should be
described by some generalization of the ${\cal N}=(2,0)$, $d=6$
tensor multiplet. \footnote{Review of this theory was given in
\cite{saemann}.} It should be noted, however, that theories that
involve antisymmetric tensor fields that interact with the
Yang-Mills field and, consequently, non-trivially transform with
respect to non-Abelian gauge groups, do exist. Such tensors can be
encountered in some gauged supergravities, such as $d=7$
supergravity \cite{PPvN} with $3-$tensor fields forming $SO(5)$
vector of internal symmetry, obtained from $d=11$ theory on $AdS_7
\times S^4$, or the $d=5$ with $2$-form fields \cite{PPvN2}
\footnote{Reviews of dimensional reduction of $d=11$ supergravity
can be found in \cite{kksugra} and \cite{kksugra2}.}, though in both
cases the shift symmetry such as \p{Babelian}, \p{Babelian3} was
gauge-fixed. \footnote{Further development of non-Abelian forms
interacting with supergravity resulted in tensor hierarchies
\cite{tenshier1,tenshier11,tenshier2,tenshier3,tenshier4}.} The earlier example of such a
construction was given by Scherk and Schwarz \cite{scherk}, where
the reduction of various theories on a group manifold \cite{dewitt}
was discussed, with a $3$-form theory among them. The resulting
action contained, in particular, massive $2$-form field charged
under the Yang-Mills gauge symmetry, along with the vector and
scalar fields and the Abelian $3$-form. It was shown in
\cite{nishino2} that the specific interactions, induced by the
dimensional reduction, lead to a consistent theory, which equations
of motion do not impose algebraic constraints on the fields. The
consistency of equations of motion makes this theory and other
analogous theories obtained by reduction of higher $p$-forms
interesting from the quantum point of view, which requires a study
of their gauge structure and possible introduction of St\"uckelberg
fields.

Although some examples of models containing interacting
antisymmetric tensor fields are known, important questions regarding
their structure still remain open. All the free bosonic and
fermionic $p$-form models are the reducible gauge theories with
various stages of reducibility. However, the aspects of reducibility
for interacting $p$-form models have never been studied. In the work under
consideration, we are going to fill this gap.

In this paper, we discuss the reducibility properties of
transformations that naturally appear in non-Abelian tensor theories
that are obtained by dimensional reduction on a group manifold. Such
manifolds contain a dimensionful parameter that is naturally related
to the interaction. At first, we describe our reduction prescription that mostly follows
\cite{scherk} starting from the free Abelian $3$-form model. We show
that gauge transformations of the fields that appear in this model
form a non-Abelian algebra and there exists a particular set of
parameters for which transformations are invariant, proving that
non-Abelian part of the transformations is the first-stage reducible. Then we extend these considerations to the system
obtained by dimensional reduction of a free $4$-form to show that
non-Abelian transformations in this case are the second-stage reducible. The results of this Section allow us to extend the
consideration to the general case to show that the system obtained
by dimensional reduction of a $p$-form model minimally coupled to
gravity contains non-Abelian antisymmetric tensor fields of rank
$p-1$ and less and the resulting transformations are $p-2$ stage
reducible.

Antisymmetrized expressions in this paper are assumed to include a
normalization coefficient that is equal to one divided by the number
of terms required for antisymmetrization. For example,
antisymmetrized product of a vector $V_{\mu_1}$ and a totally
antisymmetric tensor with $p$ indices $A_{\mu_2 \mu_3 \ldots
\mu_{p+1}}$ is \be\label{antisym1} V_{[\mu_1}A_{\mu_2 \mu_3 \ldots
\mu_{p+1}]} = \frac{1}{p+1} \big( V_{\mu_1} A_{\mu_2 \mu_3 \ldots
\mu_{p+1}} -  V_{\mu_2} A_{\mu_1 \mu_3 \ldots \mu_{p+1}} - V_{\mu_2}
A_{\mu_2 \mu_1 \ldots \mu_{p+1}} -\ldots - V_{\mu_{p+1}} A_{\mu_2
\ldots \mu_{p} \mu_{1}}\big). \ee The antisymmetrized product of
antisymmetric $2$-tensor $F_{\mu_1 \mu_2}$ and totally antisymmetric
$p$-tensor requires not $(p+2)(p+1)$ but $(p+1)(p+2)/2$ terms, and
\bea\label{antisym2} F_{[\mu_1 \mu_2 }A_{\mu_2 \mu_3 \ldots
\mu_{p+1}]} &=& \frac{2}{(p+1)(p+2)} \big( F_{\mu_1\mu_2} A_{\mu_3
\mu_4 \ldots \mu_{p+2}} -  F_{\mu_1 \mu_3 } A_{\mu_2 \mu_4 \ldots
\mu_{p+2}} -\nn
\\ &&- F_{\mu_3 \mu_2 } A_{\mu_1 \mu_4 \ldots \mu_{p+1}}-\ldots +
F_{\mu_{p+1}\mu_{p+2}} A_{\mu_3 \mu_4 \ldots \mu_{1} \mu_{2}}\big).
\eea Symbol $|$ is used to separate indices not subjected to
antisymmetrization.

\section{Dimensionally reduced $3$-form model}
\subsection{Prescriptions for dimensional reduction}
The simplest system that can produce non-Abelian antisymmetric
fields upon dimensional reduction is the $3$-form minimally coupled
to gravitation
 \be
  \label{initaction}
  S = -\frac{1}{24}\int d^D X
\sqrt{-\det G}F_{MNPQ}F^{MNPQ }-\int d^D X \sqrt{-\det G} R[G], \;\;
F_{MNPQ} = 4\partial_{[M} C_{NPQ]},
 \ee
where the metric is assumed to have ``mostly plus'' signature. It
possesses the diffeomorphism invariance and $3$-form gauge symmetry
\be\label{CLambda} \delta_{\Lambda} C_{MNP} = 3\partial_{[M}
\Lambda_{NP]}. \ee Dimensional reduction of this action on a group
manifold was briefly discussed in \cite{scherk},  where the reduced
action was derived and the gauge transformation laws of all the
fields were found, which included some non-Abelian transformations
of the tensor fields. It was noted in \cite{nishino2} that action
derived in \cite{scherk} leads to consistent equations of motion, as
divergence of each of these equations is not a forbiddingly strong
algebraic constraint on the fields, as happens in the case of
minimal coupling of the antisymmetric tensor to the Yang-Mills
field, but a combination of other equations of motion. Therefore,
this system is of interest in description of non-Abelian tensor
fields.

In \cite{scherk} the most emphasis was put on the properties of the
induced potential for the scalar fields that appear as some of
the components of the reduced metric, and properties of the tensor
fields were discussed only briefly. Let us revisit this
construction, omitting these scalar fields. This would lead to
inconsistency in the equation of motion of gravity, but, as we are
interested mostly in the properties of antisymmetric tensor fields
in the flat space, this is not an issue.

Let us perform the dimensional reduction following way. As in
\cite{dewitt}, one can split the coordinates $X^{M}$ into two
subsets, $X^M = \big( x^\mu, y^a \big)$, and parameterize the metric
as
 \be
 \label{Gstruct}
 G_{MN} = \left(
\begin{array}{c c}
g_{\mu\nu} + G_{\mu a}\big( g^{-1} \big){}^{ab} G_{\nu b} & G_{\mu a} \\
G_{b \nu }  & g_{ab}
 \end{array}
\right)
 \ee
Under such splitting, the $C_{MNP}$ produces four fields
 \be
 \label{Cfields}
 C_{\mu\nu\rho}, \;\; C_{\mu\nu c}, \;\; C_{\mu a
b}, \;\; C_{abc}.
\ee
To perform the reduction, we constrain the
metric, the antisymmetric field and the parameters $\xi^{M}$,
$\Lambda_{MN}$ by making following assumptions:
\begin{enumerate}
\item $g_{ab}$ in \p{Gstruct} is a metric on some compact Lie group ${\cal G}$ which depends on $y^{a}$ only,
with $a=1,\ldots, \mbox{dim}{\cal G}$. The size of this compact manifold is characterized by a
dimensionful parameter $\kappa$. As it should be for a group, it
possesses $\mbox{dim}{\cal G}$ Killing vectors $e^a{}_{\ba} =
e^a{}_{\ba}(y)$ ;
\item The $y$ dependence of other objects is determined by the indices they carry and can be reduced to the
Killing vector factors. For example, a  field with single group
index $q^a (X)$ is assumed to have structure $q^{a}(X) =
e^{a}{}_{\ba}(y)q_{\ba}(x)$.
\end{enumerate}
The latter rule applies to both parameters and components of the
fields. In particular, we take that
 \bea
 \label{redpost}
\xi^{\mu} = \xi^{\mu}(x), \;\; \xi^{a} = \kappa^{-1} e^a_{\ba}(y) \lambda_{\ba}(x), \;\; g_{\mu\nu} = g_{\mu\nu}(x), \;\; G_{\mu a} = G_{a \mu} =\kappa^{-1} e_{a\ba}(y) A_{\mu \ba}(x), \nn \\
C_{\mu\nu\rho}=C_{\mu\nu\rho}(x), \;\; C_{\mu\nu c} = e_{c\bc}C_{\mu\nu \bc}(x), \;\; C_{\mu a b} = e_{a\ba}e_{b\bb} C_{\mu \ba \bb}(x), \;\; C_{abc} = e_{a\ba}e_{b\bb}e_{c\bc} C_{\ba \bb \bc}(x), \\
\Lambda_{\mu\nu} = \Lambda_{\mu\nu}(x), \;\; \Lambda_{\mu a} =
e_{a\ba}\Lambda_{\mu\ba}, \;\; \;\; \Lambda_{ab} =
e_{a\ba}e_{b\bb}\Lambda_{\ba \bb}(x),\nn
 \eea
 where $e_{a\ba}\equiv
g_{ab}e^b{}_\ba$. As the metric and the Killing vectors are assumed
to be dimensionless, and the coordinates have dimension mass$^{-1}$, the parameter $\kappa$ with dimension of mass was used to canonically normalize the Yang-Mills
field.

Each group possesses the $\mbox{dim}{\cal G}$ Killing vectors, which
can be constructed explicitly using the method of nonlinear
realizations. They have two important properties:
\begin{enumerate}
\item Symmetrized covariant derivative of each of the vectors $e_{a\ba}$ is zero, $\nabla_{a}e_{b\bc} + \nabla_b e_{a\bc} =\partial_a e_{b\bc} +\partial_b e_{a\bc} +2 \Gamma_{ab}^c(g_{mn})e_{c\bc}=0$, and
\item The Lie derivative of the one along the other is again a linear combination of the Killing vectors ${\cal L}\big( e_\ba \big)e_\bb \sim f_{\ba\bb\bc}e_{\bc}$, where $f_{\ba \bb \bc}$ are the structure constants of the appropriate Lie algebra.
\end{enumerate}
We choose the following normalization of the Killing vectors:
\be\label{eprop}
e_{a\ba}e^{a}_\bb = \delta_{\ba \bb}, \;\; e_{a\ba}e_{b\ba} = g_{ab}, \;\;  \;\; \partial_{a}e_{b\bc} - \partial_{b}e_{a\bc} = -\kappa f_{\ba \bb \bc} e_{a\ba} e_{b\bb},
\ee
where again the parameter $\kappa$ was introduced to take into account the dimension of $y^a$. As the group is assumed to be compact, structure constants can be chosen completely antisymmetric and normalized as $f_{\ba\bb\bc}f_{\ba\bb\bd} = 2\delta_{\bc\bd}$. We, therefore, do not distinguish upper and lower barred indices.

Note that in \cite{scherk} the $g_{ab}$ components of the metric had structure $g_{ab}=e_{a\ba}e_{b\bb}\Phi_{\ba\bb}(x)$ and were, therefore, $x^\mu$ dependent. To simplify the construction we take $\Phi_{\ba\bb}(x) = \delta_{\ba\bb}$.

\subsection{Field redefinition and gauge transformations}
The action \p{initaction} is invariant under diffeomorphisms and
Abelian gauge transformations of $C_{MNP}$ \p{CLambda}.
Transformations of the components of $G_{MN}$ and $C_{MNP}$ at a
point, induced by diffeomorphism $\delta X^{M} =- \xi^{M}(X^N)$,
read
 \bea
 \label{deltadif}
\delta_\xi G_{MN} &=& G\pr_{MN}(X)-G_{MN}(X)=  {\cal L}\big(\xi^P \big)G_{MN} = \xi^P \partial_P G_{MN} + \partial_M \xi^P \, G_{PN} +  \partial_N \xi^P \, G_{MP}, \nn \\
\delta_\xi C_{MNP} &=& C\pr_{MNP}(X)-C_{MNP}(X)= {\cal L}\big(\xi^P \big)C_{MNP} = \\
&=& \xi^Q \partial_Q C_{MNP} + \partial_M \xi^Q \, C_{QNP} +
\partial_N \xi^Q \, C_{MQP} +\partial_P \xi^Q \, C_{MNQ},\nn
 \eea
where ${\cal L}\big(\xi^P\big)$ is the Lie derivative along the
vector $\xi^P$. It should be noted that the assumptions made above
are fully compatible with the transformation laws \p{deltadif},
\p{CLambda}. As we consider transformations at a point in $X$ space,
$e_{a\ba}$, being fixed functions of coordinates, are not varied.
Therefore, for example, $\delta C_{\mu\nu a}(x,y) =
e_{a\ba}(y)\delta C_{\mu\nu \ba}(x)$, and $y^a$ dependence of left
hand side of each of transformation laws \p{deltadif}, \p{CLambda}
reduces to appropriate $e_{a\ba}e_{b\bb}\ldots$ factors. After
substitution of \p{redpost} into the right hand side of \p{deltadif},
\p{CLambda} and simplification involving \p{eprop} similar
$e_{a\ba}e_{b\bb}\ldots$ factors appear, and it becomes possible to
reduce \p{deltadif}, \p{CLambda} to transformations of
$x^\mu$-dependent fields. In particular, transformation of $G_{ab} =
g_{ab}(y)$ vanishes exactly if \p{eprop} hold, even though parameter
$\lambda_{\ba}$ depends on $x^\mu$. Transformation of $G_{\mu a}$
confirms that $A_{\mu \ba}$ is the Yang-Mills field:
 \be
 \label{Aprop}
 \delta A_{\mu\ba} = {\cal L}\big( \xi^\sigma
\big)A_{\mu\ba} + \partial_\mu \lambda_\ba +
f_{\ba\bm\bn}\lambda_\bm A_{\mu \bn}.
 \ee
Using \p{Aprop}, one can show that $\delta g_{\mu\nu}$ reduces to
standard diffeomorphism
 \be\label{gmunuprop} \delta g_{\mu\nu} ={\cal L}\big( \xi^\sigma
\big)g_{\mu\nu},
 \ee
if $g_{\mu\nu}$ does not depend on $y^a$. In further considerations,
we consider it flat, $g_{\mu\nu} = \eta_{\mu\nu}$. Then
diffeomorphisms reduce to Poincar\'e transformations and can be
discarded.

Transformations of $C_{MNP}$ fields reduce similarly:
 \bea
 \label{Ctrans}
\delta C_{\mu\nu\rho} &=& 3 \kappa^{-1}  \partial_{[\mu} \lambda_{\ba}\, C_{\nu\rho]\ba}+ 3\partial_{[\mu}\Lambda_{\nu\rho]} ,  \\
\delta C_{\mu\nu\bc} &=&  2\kappa^{-1}  \partial_{[\mu} \lambda_{\bd} C_{\nu]\bc \bd}  + f_{\bc \bm \bn} \lambda_{\bm} C_{\mu\nu \bn} + 2\partial_{[\mu} \Lambda_{\nu] \bc}, \nn \\
\delta C_{\mu\ba\bb} &=&  \kappa^{-1} \partial_\mu \lambda_{\bd} C_{\ba\bb \bd}  +2 f_{[\ba|\bm\bn}\lambda_\bm C_{\mu\bn|\bb]}  +\kappa f_{\ba\bb\bc}\Lambda_{\mu \bc} + \partial_{\mu}\Lambda_{\ba\bb}, \nn \\
\delta C_{\ba\bb\bc} &=& 3f_{[\ba|\bm\bn}\lambda_\bm C_{\bn|\bb\bc]}  -
3\kappa f_{[\ba\bb|\bd}\Lambda_{\bd|\bc]}.\nn
 \eea
This basis of fields, as was noted in \cite{scherk}, is inconvenient, as the fields do not transform
homogeneously with respect to the Yang-Mills group. The suitable fields are
 \bea
 \label{Credef}
B_{\mu\nu\rho}&=& C_{\mu\nu\rho} - 3\kappa^{-1} A_{[\mu \ba} C_{\nu\rho] \ba} + 3\kappa^{-2} C_{[\mu\ba\bb}A_{\nu\ba}A_{\rho]\bb} - \kappa^{-3} C_{\ba\bb\bc}A_{\mu\ba}A_{\nu\bb}A_{\rho\bc}, \nn \\
B_{\mu\nu\ba} &=& C_{\mu\nu\ba} -2 \kappa^{-1} A_{[\mu\bb}C_{\nu]\ba\bb} + \kappa^{-2}C_{\ba\bb\bc}A_{\mu\bb}A_{\nu\bc}, \nn \\
B_{\mu\ba\bb} &=& C_{\mu\ba\bb} - \kappa^{-1}A_{\mu\bc}C_{\ba\bb\bc}, \nn \\
B_{\ba\bb\bc} &=& C_{\ba\bb\bc}.
 \eea
It is worth noting that there is a convenient way to obtain these relations. The $D$-bein of the metric
and its inverse can be presented as \cite{scherk}
 \be
 \label{Dbein}
E_{M}{}^{{\bar M}} = \left(
 \begin{array}{c c}
\delta_\mu{}^{{\bar\mu}} & \kappa^{-1}A_{\mu \ba} \\
0  & e_{a\ba}
\end{array}
\right), \;\; \big( E^{-1} \big)_{{\bar M} }{}^M =  \left(
 \begin{array}{c c}
\delta_{\bar\mu}{}^{\mu} & -\kappa^{-1} A_{\bar\mu \ba}e^{a}{}_\ba \\
0  & e^a{}_{\ba}
\end{array}
\right).
 \ee
Then the $B$ fields can be extracted as projections from the relation
 \be
 \label{Bnewdef}
B_{{\bar M}{\bar N}{\bar P}} = \big( E^{-1} \big)_{{\bar M} }{}^M \big( E^{-1} \big)_{{\bar N} }{}^N \big( E^{-1} \big)_{{\bar P} }{}^P C_{MNP}.
\ee
Note that the in right hand side of \p{Bnewdef} formally contains $y$-dependent terms, but they
mutually cancel. Also, as we are working in the flat space, $\mu$ and $\bar\mu$ indices can be
identified.

The transformation laws of $B_{\mu\nu\rho}$, $B_{\mu\nu\ba}$, $B_{\mu\ba\bb}$, $B_{\ba\bb\bc}$ read
\bea
\label{Btrans}
\delta B_{\mu\nu\rho} &=& 3\partial_{[\mu}{\widetilde \Lambda}_{\nu\rho]} +\partial_{\rho}{\widetilde \Lambda}_{\mu\nu} - 3\kappa^{-1} \cF_{[\mu\nu\ba}{\widetilde \Lambda}_{\rho]\ba}, \nn \\
\delta B_{\mu\nu\ba} &=& 2\nabla_{[\mu} {\widetilde \Lambda}_{\nu]\ba} + f_{\ba\bm\bn}\lambda_\bm B_{\mu\nu\bn}, \nn \\
\delta B_{\mu\ba\bb} &=& \nabla_\mu {\widetilde \Lambda}_{\ba\bb} +\kappa f_{\ba\bb\bc}{\widetilde \Lambda}_{\mu\bc} + 2f_{[\ba|\bm\bn}\lambda_\bm B_{\mu\bn|\bb]}, \\
\delta B_{\ba\bb\bc} &=& 3 f_{[\ba|\bm\bn}\lambda_\bm
B_{\bn|\bb\bc]} - \kappa f_{[\ba\bb|\bd}{\widetilde
\Lambda}_{\bd|\bc]},\nn
 \eea
 where the new parameters, covariant
derivative and $A_{\mu\ba}$ field strength are defined as
 \bea\label{Bparams}
{\widetilde \Lambda}_{\mu\nu} &=& {\Lambda}_{\mu\nu} + 2\kappa^{-1}A_{[\mu\ba} \Lambda_{\nu]\ba}  + \kappa^{-2}\Lambda_{\ba\bb}A_{\mu\ba}A_{\nu\bb}, \nn \\
{\widetilde \Lambda}_{\mu \ba} &=& \Lambda_{\mu\ba} + \kappa^{-1}A_{\mu\bb}\Lambda_{\ba\bb}, \;\; {\widetilde \Lambda}_{\ba \bb} = \Lambda_{\ba \bb}, \nn \\
\nabla_\mu q_{\ba} &=& \partial_\mu q_\ba - f_{\ba\bm\bn}A_{\mu\bm}q_{\bn}, \;\; \big[ \nabla_\mu, \nabla_\nu  \big]q_\ba = -f_{\ba\bm\bn}\cF_{\mu\nu\bm}q_\bn, \nn \\
\cF_{\mu\nu\ba} &=& \partial_\mu A_{\nu\ba} - \partial_\nu A_{\mu\ba} - f_{\ba\bm\bn}A_{\mu\bm}A_{\nu\bn}.
 \eea
Like the $B$ fields, the ${\widetilde\Lambda}$ parameters can be
obtained from the relation ${\widetilde\Lambda}_{{\bar M} {\bar N}}
=  \big( E^{-1} \big)_{{\bar M} }{}^M \big( E^{-1} \big)_{{\bar N}
}{}^N \Lambda_{MN}$.

Together with the Yang-Mills field transformations
 \be\label{deltaAmu}
\delta A_{\mu\ba} = \partial_\mu \lambda_{\ba} + f_{\ba\bb\bc}\lambda_{\bb}A_{\mu\bc}
 \ee
the laws \p{Btrans} form a complete set of gauge symmetries of the system.

\subsection{Algebra of gauge transformations}
It can be verified that the algebra of gauge transformations \p{Btrans}, \p{deltaAmu} is not Abelian.
This is obvious for the Yang-Mills field, but less so for others. Let $\delta_i$ be the transformation
with the set of parameters
 \be\label{deltai}
\big\{ \Lambda_i  \big\} =\big\{ \lambda_{i\ba}, {\widetilde \Lambda}_{i\mu\nu}, {\widetilde \Lambda}_{i\mu\ba},  {\widetilde \Lambda}_{i\ba\bb} \big\}.
 \ee

Then $\delta_1 B_{\mu\nu\ba}$ would read
 \be\label{deltaBmunu1}
\delta_1 B_{\mu\nu\ba} = \nabla_\mu {\widetilde \Lambda}_{1\nu\ba}-
\nabla_\nu {\widetilde \Lambda}_{1\mu\ba} -\kappa^{-1}\cF_{\mu\nu\bb}{\widetilde \Lambda}_{1\ba\bb} +
f_{\ba\bm\bn}\lambda_{1\bm} B_{\mu\nu\bn}.
 \ee
Transforming \p{deltaBmunu1} with parameters $\big\{ \Lambda_2 \big\}$, one can simplify the result to
 \bea\label{deltaBmunu21}
\delta_2\delta_1 B_{\mu\nu\ba} &=& - f_{\ba\bm\bn}\nabla_\mu \lambda_{2\bm}\, {\widetilde\Lambda}_{1\nu\bn} + f_{\ba\bm\bn}\nabla_\nu \lambda_{2\bm}\, {\widetilde\Lambda}_{1\mu\bn} +f_{\ba\bm\bn} \lambda_{1\bm}\big( \nabla_{\mu} {\widetilde\Lambda}_{2\nu\bn}  -\nabla_{\nu} {\widetilde\Lambda}_{2\mu\bn}  \big)-  \\&& - \kappa^{-1} \cF_{\mu\nu \bn} f_{\bb\bm\bn}\lambda_{2\bm}{\widetilde\Lambda}_{1\ba\bb} - \kappa^{-1} f_{\ba\bm\bn} \lambda_{1\bm} \Lambda_{2\bn\bp} \cF_{\mu\nu\bp} + f_{\ba\bm\bn}f_{\bn\bp\bq}\lambda_{1\bn}\lambda_{2\bp} B_{\mu\nu\bq}.\nn
 \eea
Swapping indices $1\leftrightarrow 2$ and subtracting the result, one can collect terms with
$B_{\mu\nu\ba}$, $\cF_{\mu\nu\ba}$ and the remaining ones. Therefore, with some use of Jacobi
identity, commutator
 \bea\label{deltaBmunucom}
\big[\delta_2,\delta_1 \big] B_{\mu\nu\ba} &=& \nabla_{\mu} \big( f_{\ba\bm\bn} \lambda_{1\bm} {\widetilde \Lambda}_{2\nu\bn}-f_{\ba\bm\bn} \lambda_{2\bm} {\widetilde \Lambda}_{1\nu\bn} \big) - \nabla_{\nu} \big( f_{\ba\bm\bn} \lambda_{1\bm} {\widetilde \Lambda}_{2\mu\bn}-f_{\ba\bm\bn} \lambda_{2\bm} {\widetilde \Lambda}_{1\mu\bn} \big) - \nn \\
&&-\kappa^{-1} \cF_{\mu\nu\bb}\left( f_{\ba\bm\bn}\big( \lambda_{1\bm}{\widetilde \Lambda}_{2 \bn\bb} -  \lambda_{2\bm}{\widetilde \Lambda}_{1 \bn\bb} \big) + f_{\bb\bm\bn}\big( \lambda_{1\bm}{\widetilde \Lambda}_{2 \ba\bn} -  \lambda_{2\bm}{\widetilde \Lambda}_{1 \ba\bn} \big) \right)+\nn \\ &&+ f_{\ba\bn\bq}f_{\bn\bm\bp}\lambda_{1\bm}\lambda_{2\bp} B_{\mu\nu\bq}
 \eea
can be written as a $B_{\mu\nu\ba}$ transformation with parameters
 \bea
 \label{commdelta}
\lambda_{3\ba} &=& f_{\ba\bb\bc}\lambda_{1\bb}\lambda_{2\bc}, \nn \\
{\widetilde \Lambda}_{3\mu\bc} &=& f_{\bc\bm\bn} \big( \lambda_{1\bm}{\widetilde \Lambda}_{2\mu\bn} - \lambda_{2\bm}{\widetilde \Lambda}_{1\mu\bn}\big), \\
{\widetilde \Lambda}_{3\ba\bb} &=& f_{\ba\bm\bn} \big( \lambda_{1\bm}{\widetilde \Lambda}_{2\bn\bb} - \lambda_{2\bm}{\widetilde \Lambda}_{1\bn\bb}    \big) - f_{\bb\bm\bn} \big( \lambda_{1\bm}{\widetilde \Lambda}_{2\bn\ba} - \lambda_{2\bm}{\widetilde \Lambda}_{1\bn\ba}    \big).\nn
 \eea

Relations \p{commdelta} are, actually, all the nontrivial relations
of the gauge algebra. Commutator $\big[\delta_2,\delta_1 \big]$,
acting on $B_{\mu\ba\bb}$, $B_{\ba\bb\bc}$, $B_{\mu\nu\rho}$ and
$A_{\mu\ba}$, results in transformations with parameters given by
\p{commdelta}. The ${\widetilde\Lambda}_{\mu\nu}$ transformation
commutes with all others and does not appear at the right hand side
of commutators, and whole algebra of transformations is a sum of
Abelian ${\widetilde \Lambda}_{\mu\nu}$ and non-Abelian $\big\{
\lambda_{\ba}, {\widetilde \Lambda}_{\mu\ba},  {\widetilde
\Lambda}_{\ba\bb} \big\}$ subalgebras.

Relations \p{commdelta} prove that we are dealing with the
non-Abelian theory, with non-trivially commuting transformations of
antisymmetric tensor fields. Remarkably, the algebra closes without
reference to the equations of motion.

\subsection{The reduced action}
Due to their simpler transformation laws, it is natural to formulate the reduced action in terms of $B$ fields. As an intermediate step, it is useful to derive field strengths which transformations, like \p{Btrans}, do not involve derivatives of $\lambda_\ba$. They should be combinations of $C$ field strengths with powers of $A_{\mu\ba}$.

The field strength $F_{MNPQ}$ after factorization of of $y^a$ dependence due to \p{eprop} produces the following objects
\bea\label{Fsplit}
F_{\mu\nu\rho\sigma} &=& 4\partial_{[\mu} C_{\nu\rho\sigma]}, \nn \\
F_{\mu\nu\rho a} &=& e_{a\ba}F_{\mu\nu\rho \ba}, \;\; F_{\mu\nu\rho \ba} =3\partial_{[\mu} C_{\nu\rho]\ba}, \nn \\
F_{\mu\nu a b} &=& e_{a\ba} e_{b\bb} F_{\mu\nu\ba \bb}, \;\; F_{\mu\nu\ba \bb} = \partial_{[\mu} C_{\nu] \ba \bb} - \kappa f_{\ba\bb\bc}C_{\mu\nu\bc}, \nn \\
F_{\mu abc} &=& e_{a\ba} e_{b\bb} e_{c\bc} F_{\mu\ba \bb \bc}, \;\; F_{\mu\ba \bb \bc} = \partial_{\mu}C_{\ba\bb\bc} + 3\kappa f_{[\ba\bb|\bd}C_{\mu\bd|\bc]} , \nn \\
F_{abcd} &=& e_{a\ba} e_{b\bb} e_{c\bc} e_{d\bd} F_{\ba \bb \bc \bd}, \;\; F_{\ba \bb \bc \bd} = -6\kappa f_{[\ba \bb |\bm} C_{\bm |\bc \bd]},\nn
 \eea
which transform non-homogeneously with respect to the Yang-Mills symmetry. The correct field strengths can be found from the analog of the formula \p{Bnewdef}
 \be\label{Hdef}
H_{{\bar M}{\bar N}{\bar P}{\bar Q}} = \big( E^{-1} \big)_{{\bar M} }{}^M \big( E^{-1} \big)_{{\bar N} }{}^N \big( E^{-1} \big)_{{\bar P} }{}^P \big( E^{-1} \big)_{{\bar Q} }{}^Q F_{MNPQ}.
 \ee
They can be expressed in terms of $B$ fields as
 \bea\label{HB}
H_{\mu\nu\rho\sigma} &=& 4\partial_{[\mu}B_{\nu\rho\sigma]} + 6 \kappa^{-1} \cF_{[\mu\nu\ba}B_{\rho\sigma]\ba} , \nn \\
H_{\mu\nu\rho\ba} &=& 3 \nabla_{[\mu} B_{\nu\rho]\ba} +3 \kappa^{-1}\cF_{[\mu\nu\bb}B_{\rho]\ba\bb} , \nn \\
H_{\mu\nu\ba\bb} &=& 2\nabla_{[\mu} B_{\nu]\ba\bb} + \kappa^{-1} \cF_{\mu\nu\bc}B_{\ba\bb\bc} - \kappa f_{\ba\bb\bc}B_{\mu\nu\bc},  \\
H_{\mu\ba\bb\bc} &=& \nabla_\mu B_{\ba\bb\bc} +3\kappa f_{[\ba\bb|\bd} B_{\mu\bd|\bc]} , \nn \\
H_{\ba\bb\bc\bd} &=& -6\kappa f_{[\ba\bb|\bm}B_{\bm|\bc\bd]}. \nn
 \eea

Due to relation \p{Hdef} the $F_{MNPQ}F^{MNPQ}$ in the action
\p{initaction} reduces to the sum of various $H$ field strengths.
Also noting that, as a result of the specific parametrization of the
metric,
 \be\label{Gdetrel}
\det G_{MN} = \det g_{\mu\nu} \, \det g_{ab} = -\det g_{ab}.
 \ee
Therefore, the integration over $y$ produces a factor $\sim
\kappa^{-{\mbox{dim}}{\cal G}}$ (volume of the compact subspace).
Ignoring this factor and the cosmological term, one can write down
the action as an integral over $x^\mu$:
 \bea\label{action2}
S &=& -\frac{1}{24} \int d^n x \left(  H_{\mu\nu\rho\sigma}H^{\mu\nu\rho\sigma} +4 H_{\mu\nu\rho\bm}H^{\mu\nu\rho}{}_{\bm} + 6 H_{\mu\nu \bc\bd}H^{\mu\nu}{}_{\bc\bd} + \right.\nn \\
&& \left. +4 H_{\mu\ba\bb\bc}H^\mu{}_{\ba \bb \bc} + H_{\ba\bb\bc\bd}H_{\ba\bb\bc\bd} +6 \cF_{\mu\nu\ba}\cF^{\mu\nu}{}_\ba  \right).
 \eea
Here, the Yang-Mills Lagrangian is the only nontrivial contribution
that comes from the gravitational part of the action \cite{dewitt}.

The equations of motion that follow from the action \p{action2} were
studied in \cite{nishino2} and were shown to be consistent,
implying that \p{action2} is a sensible non-Abelian tensor model.

\subsection{Reducibility of gauge transformations}
One of the known peculiarities of the Abelian antisymmetric tensor models, that
distinguishes them from the conventional gauge theories, is the fact
that their gauge transformations are reducible. This means that
different gauge parameters generate the same transformation of the
fields, or, equivalently, there exists a non-trivial set of
parameters for which the gauge fields do not transform at all. These
properties are to be expected from interacting models also if they
possess standard kinetic terms for $p$-form field. If the gauge
symmetries of a free theory are broken by the interactions, one should restore them by adding specific auxiliary fields. In
particular, reducibilty should be the property of the transformation laws
\p{Btrans}, \p{deltaAmu}, which are the gauge symmetries of the
dimensionally reduced $3$-form on a curved background, if this model
actually provides a correct description of non-Abelian tensor
fields. To show that this is the case, one should find a parameter
or a combination of parameters for which both \p{Btrans} and
\p{deltaAmu} are exactly zero.

In search of this set of parameters, it is important to note that
the transformation of the Yang-Mills field \p{deltaAmu} involves
only one parameter $\lambda_\ba$ and can not be zero for
$\lambda_\ba\neq 0$ if  $A_{\mu\ba}$ is not a pure gauge, as
equation
 \be\label{deltaAmured}
\delta A_{\mu \ba} =\partial_\mu \lambda_{\ba} + f_{\ba\bb\bc}\lambda_{\bb}A_{\mu\bc} =
\nabla_\mu \lambda_\ba =0
 \ee
implies $\big[\nabla_\mu,\nabla_\nu \big]\lambda_\ba =-f_{\ba\bb\bc}\cF_{\mu\nu\bb}\lambda_{\bc}=0$. As in our case $A_{\mu\ba}$ is a dynamical field, any transformation with $\lambda_\ba \neq 0$ is nontrivial. Therefore, only the transformations with $\lambda_\ba =0$ are of interest in the context of reducibility.

The $B_{\mu\nu\ba}$ transformation \p{Btrans} with $\lambda_\bm =0$ reads
 \be\label{B2trans}
\delta B_{\mu\nu\ba} = \nabla_\mu {\widetilde \Lambda}_{\nu\ba}-\nabla_\nu {\widetilde \Lambda}_{\mu\ba}
 -\kappa^{-1}\cF_{\mu\nu\bb}{\widetilde \Lambda}_{\ba\bb}.
 \ee
Generalizing the Abelian case, one can expect that ${\widetilde \Lambda}_{\mu\ba}= {\widetilde \Lambda}^{triv}_{\mu\ba}=
\nabla_{\mu}\omega_{\ba}$ could lead to a trivial transformation. Though
 \be\label{dLambda}
\nabla_\mu {\widetilde \Lambda}^{triv}_{\nu\ba}  - \nabla_\nu {\widetilde \Lambda}^{triv}_{\mu\ba} =
\big[ \nabla_\mu, \nabla_{\nu}]\omega_{\ba} = -f_{\ba\bb\bc}\cF_{\mu\nu\bb}\omega_\bc \neq 0,
 \ee
this can be compensated by choosing ${\widetilde\Lambda}_{\ba\bb}={\widetilde\Lambda}^{triv}_{\ba\bb} = -\kappa f_{\ba\bb\bc}\omega_\bc$.
Exactly the same parameters nullify $B_{\mu\nu\ba}$ and $B_{\ba\bb\bc}$ transformations
 \bea\label{B10trans}
\delta B_{\mu\ba\bb} &=& \nabla_\mu {\widetilde \Lambda}_{\ba\bb} +\kappa f_{\ba\bb\bc}{\widetilde \Lambda}_{\mu\bc}, \nn \\
\delta B_{\ba\bb\bc} &=&  - \kappa f_{\ba\bb\bd}{\widetilde \Lambda}_{\bd\bc} + \kappa f_{\ba\bc\bd}{\widetilde \Lambda}_{\bd\bb} - \kappa f_{\bb\bc\bd}{\widetilde \Lambda}_{\bd\ba},
 \eea
the latter due to Jacobi identity. The trivial transformation of $B_{\mu\nu\rho}$ can be found if one,
along with ${\widetilde \Lambda}_{\mu\ba} = \nabla_{\mu}\omega_{\ba}$, sets ${\widetilde \Lambda}_{\mu\nu} = \kappa^{-1} \cF_{\mu\nu\ba}\omega_\ba$ as a compensation. Keeping in mind what has been said on the reducibility of the $3$-form models, one can note that the trivial transformations are defined by the parameters
 \be\label{trivtr}
\lambda^{triv}_\ba=0, \;\; {\widetilde \Lambda}^{triv}_{\mu\ba} = \nabla_{\mu}\omega_{\ba}, \;\;
{\widetilde\Lambda}^{triv}_{\ba\bb} = -\kappa f_{\ba\bb\bc}\omega_\bc, \;\;
{\widetilde \Lambda}^{triv}_{\mu\nu} = \kappa^{-1} \cF_{\mu\nu\ba}\omega_\ba +
\partial_{\mu}\zeta_{\nu} - \partial_\nu \zeta_\mu, \;\; \zeta_\mu \sim \zeta_\mu + \partial_\mu \zeta.
 \ee
Therefore, the transformations of non-Abelian fields are reducible,
just as in the Abelian case, and the transformation of the $3$-form
is of second stage of reducibility.

As has been noted in the Introduction, the existence of trivial transformations $\Lambda^{triv}(\omega)$ implies that the different gauge parameters $\Lambda$ and $\Lambda^\prime$ should lead to the similar transformation if $\Lambda^\prime = \Lambda+ \Lambda^{triv}(\omega)$, which means that the gauge parameters are defined up to their own gauge transformation. In the case of a system obtained by the dimensional reduction of the $3$-form coupled to gravity, these transformations of
${\widetilde\Lambda}_{\mu\nu}$, ${\widetilde \Lambda}_{\mu\ba}$, ${\widetilde\Lambda}_{\ba\bb}$ are
\bea\label{deltaLambda}
{\widetilde\Lambda}^{\prime}_{\mu\nu} &=& {\widetilde\Lambda}_{\mu\nu} + \kappa^{-1} \cF_{\mu\nu\ba}\omega_\ba +
\partial_{\mu}\zeta_{\nu} - \partial_\nu \zeta_\mu, \nn \\
{\widetilde\Lambda}^{\prime}_{\mu\ba} &=& {\widetilde\Lambda}_{\mu\ba} +  \nabla_{\mu}\omega_{\ba}, \\
{\widetilde\Lambda}^{\prime}_{\ba\bb} &=& {\widetilde\Lambda}_{\ba\bb}  -\kappa f_{\ba\bb\bc}\omega_\bc,\nn
\eea
where the parameter of the reducible Abelian transformation $\zeta_\mu$ is on its own defined up to a transformation $\zeta^\prime_\mu = \zeta_\mu + \partial_\mu \zeta$.

It is worth noting that \p{trivtr} are the only trivial transformations of the system, at least if the
underlying group is semi-simple, like $SU(N)$ or $SO(N)$. In this case the structure constants can be
chosen to satisfy $f_{\ba\bb\bc}f_{\ba\bb\bd} = 2\delta_{\bc\bd}$. Therefore, multiplying
$\delta B_{\ba\bb\bc}$ \p{B10trans} by $f_{\ba\bb\bd}$ and using Jacobi identity, one can reduce equation $\delta B_{\ba\bb\bc}=0$ to
 \be\label{deltaB0rev}
-2{\widetilde\Lambda}_{\bc\bd} + f_{\bc\bd\ba}\, f_{\ba\bm\bn}{\widetilde\Lambda}_{\bm\bn}=0, \;\; \mbox{or} \;\; {\widetilde\Lambda}_{\bc\bd}\sim f_{\bc\bd\bm}\omega_\bm.
 \ee
Similarly, there is no additional freedom for ${\widetilde \Lambda}_{\mu\bc}$, as
$f_{\ba\bb\bc}{\widetilde \Lambda}_{\mu\bc}=0$ demands ${\widetilde \Lambda}_{\mu\bc}=0$ if
$f_{\ba\bb\bc}f_{\ba\bb\bd} = 2\delta_{\bc\bd}$.

From these results one can conclude that the gauge transformations
of non-Abelian antisymmetric tensors in a system obtained by
dimensional reduction of a $3$-form coupled to gravity have the same
reducibility stages as the free Abelian ones in spite of presence of
mass terms and interactions.

\subsection{St\"uckelberg fields}
Analyzing the structure of the action \p{action2}, one can note that due to specific terms in the field strengths \p{HB}, proportional to the coupling constant $\kappa$, some of the fields acquire masses $\sim \kappa$. In particular, this happens to the non-Abelian tensor $B_{\mu\nu\ba}$, as $H_{\mu\nu\ba\bb}$ contains a term $-\kappa f_{\ba\bb\bc}B_{\mu\nu\bc}$ which adds to the Lagrangian a term $+12\kappa^2 B_{\mu\nu\ba}B^{\mu\nu}{}_{\ba}$, as we assume that $f_{\ba\bb\bc}f_{\ba\bb\bd} = 2\delta_{\bc\bd}$. If such term was added to the Abelian field, it would have broken the standard tensor gauge symmetry $\delta B_{\mu\nu} = \partial_\mu \Lambda_\nu -\partial_\mu \Lambda_\mu$. One can expect, therefore, that some of the fields present in the system are St\"uckelberg ones that effectively restore this symmetry. To show this, one should define fields those transformation laws explicitly include the shift by the parameters ${\widetilde\Lambda}_{\mu\ba}$ and ${\widetilde\Lambda}_{\ba\bb}$. The $B_{\mu\ba\bb}$ transformation law \p{Btrans} suggests that in the former case the right field is
\be\label{deltaCmu}
S_{\mu\ba} = \frac{1}{2} f_{\ba\bb\bc} B_{\mu\bb\bc}, \;\; \delta S_{\mu\ba} =  \kappa {\widetilde \Lambda}_{\mu\ba}+\nabla_\mu \gamma_\ba  + f_{\ba\bm\bn} \lambda_\bm S_{\mu\bn}, \;\; \gamma_\ba \equiv \frac{1}{2}f_{\ba\bb\bc}{\widetilde \Lambda}_{\bb\bc},
\ee
where again $f_{\ba\bb\bc}f_{\ba\bb\bd} = 2\delta_{\bc\bd}$.

Similarly, one can try to find a scalar field that shifts due to ${\widetilde \Lambda }_{\ba\bb}$ transformation. Multiplying $\delta B_{\ba\bb\bc}$ by a structure constant, one can find from \p{Btrans} that $S_{\ba\bb}\equiv B_{[\ba|\bm\bn} f_{\bm\bn|\bb]}$ transforms as
\be\label{deltaCab}
\delta S_{\ba\bb} = 2\kappa {\widehat \Lambda}_{\ba\bb} + f_{\ba\bm\bn} \lambda_\bm S_{\bn\bb} + f_{\bb\bm\bn} \lambda_\bm S_{\ba\bn}, \;\; {\widehat \Lambda}_{\ba\bb} \equiv {\widetilde \Lambda}_{\ba\bb} - \frac{1}{2}f_{\ba\bb\bc}f_{\bc\bm\bn}{\widetilde \Lambda}_{\bm\bn}.
\ee
It should be noted, however, that identically
\be\label{Cabprop}
f_{\ba\bb\bc}S_{\ba\bb}=0, \;\; f_{\ba\bb\bc} {\widehat \Lambda}_{\ba\bb}=0.
\ee
This implies that the parameter ${\widetilde\Lambda}_{\ba\bb}$ splits into $ {\widehat \Lambda}_{\ba\bb}$ and $f_{\ba\bb\bc}$-proportional part, which happens to be $\gamma_\bc$:
\be\label{Lambdaabsplit}
{\widetilde\Lambda}_{\ba\bb} = {\widehat \Lambda}_{\ba\bb} +f_{\ba\bb\bc}\gamma_\bc.
\ee
Note that due to \p{Cabprop} there is no St\"uckelberg field for $\gamma_\bc$. The difference compared to the $B_{\mu\nu\ba}$ is related to the structure of the mass term for $B_{\mu\ba\bb}$, that comes from $3\kappa f_{[\ba\bb|\bd} B_{\mu\bd|\bc]}$ term in $H_{\mu\ba\bb\bc}$, the ``field strength'' of scalar fields. This mass term can be brought to the form
\be\label{vectmass}
24\kappa^2 B_{\mu\ba\bb}B^\mu{}_{\bc\bd} Q_{\ba\bb,\bc\bd}, \;\; Q_{\ba\bb,\bc\bd} = \delta_{\ba[\bc}\delta_{\bd]\bb} - \frac{1}{2}f_{\bm\ba\bb}f_{\bm\bc\bd}.
\ee
As $Q_{\ba\bb,\bc\bd}$ is a projector, $Q_{\ba\bb,\bc\bd}Q_{\bc\bd,\bm\bn} = Q_{\ba\bb,\bm\bn}$ and $f_{\ba\bb\bc}Q_{\ba\bb,\bm\bn}=0$, only the vectors satisfying $B_{\mu\ba\bb}f_{\ba\bb\bc}=0$ acquire masses, matching the properties and number of St\"uckelberg scalars $S_{\ba\bb}$. As $f_{\ba\bb\bc}$-trace part of $B_{\mu\ba\bb}$ is the St\"uckelberg field $S_{\mu\bc}$ \p{deltaCmu}, one can conclude that $S_{\ba\bb}$ is a St\"uckelberg field that restores gauge symmetry of vector fields.

Presence of the St\"uckelberg fields also explains existence of a theory of tensor fields with non-Abelian gauge symmetry. While the considerations on the structure of $p$-form interactions \cite{henneaux1}, \cite{bekaert1} involved systems with deformations of standard Abelian symmetry $\delta B_{\mu_1 \ldots \mu_p} = (p+1)\partial_{[\mu_1} \Lambda_{\mu_2 \ldots \mu_p]}$, in the system obtained by dimensional reduction of a $3$-form mass terms explicitly break this symmetry, and the terms with derivatives of $\Lambda_{\mu\ba}$ can be present in \p{Btrans} due to the compensating St\"uckelberg fields.

\section{Reducibility in the $4$-form system}
Expecting that the dimensional reduction naturally generates a system of
non-Abelian tensors with appropriately coupled auxiliary
St\"uckelberg fields, one can study the next simplest example, the
model obtained by dimensional reduction of a free $4$-tensor field.
The reduction can be performed using similar prescriptions as in the
$3$-form case, using analogs of the formulae \p{Bnewdef} and \p{Hdef}. Skipping
unnecessary details, one can find that the reduction results in a
system of the Yang-Mills field $A_{\mu\ba}$, antisymmetric tensors
$B_{\mu\nu\rho\sigma}$, $B_{\mu\nu\rho\ba}$, $B_{\mu\nu\ba\bb}$, a
vector $B_{\mu\ba\bb\bc}$ and a scalar $B_{\ba\bb\bc\bd}$. Their
gauge transformation laws, in addition to the Yang-Mills one
\p{deltaAmu}, are
 \bea\label{B4trans}
\delta B_{\mu\nu\rho\sigma} &=& 4 \partial_{[\mu} \Lambda_{\nu \rho\sigma ]} + 6 \kappa^{-1} \cF_{[\mu \nu \bb }\Lambda_{\rho\sigma]\bb}, \nn \\
\delta B_{\mu\nu\rho\ba} &= & 3 \nabla_{[\mu} \Lambda_{\nu\rho]\ba} +3 \kappa^{-1} \cF_{[\mu\nu\bb}\Lambda_{\rho]\ba\bb} + f_{\ba\bm\bn}\lambda_{\bm} B_{\mu\nu\rho\bn}, \nn \\
\delta B_{\mu\nu\ba\bb} &=& 2 \nabla_{[\mu} \Lambda_{\nu] \ba\bb} +\kappa^{-1}\cF_{\mu\nu\bc}\Lambda_{\ba\bb\bc} - \kappa f_{\ba\bb\bc}\Lambda_{\mu\nu\bc} + 2 f_{[\ba|\bm\bn} \lambda_{\bm} B_{\mu\nu\bn|\bb]}, \\
\delta B_{\mu\ba\bb\bc} &=& \nabla_\mu \Lambda_{\ba\bb\bc} + 3\kappa f_{[\ba\bb|\bd}\Lambda_{\mu \bd|\bc]} +3 f_{[\ba|\bm\bn}\lambda_{\bm}B_{\mu\bn|\bb\bc]}, \nn \\
\delta B_{\ba\bb\bc\bd} &=& -6 \kappa f_{[\ba\bb|\bm}\Lambda_{\bm|\bc\bd]} +4 f_{[\ba|\bm\bn}\lambda_{\bm}B_{\bn|\bb\bc\bd]}.\nn
 \eea
The transformations \p{B4trans} have similar structure to \p{Btrans} and are, with exception of $\Lambda_{\mu\nu\rho}$, non-Abelian. The commutator of transformations $\delta_{1,2}$ with parameters $\big\{ \lambda_{i\ba}, \Lambda_{i\mu\nu\ba}, \Lambda_{i\mu\ba\bb}, \Lambda_{i\ba\bb\bc} \big\}$ is a transformation $\delta_3$ with
 \bea\label{B4transcom}
\lambda_{3\ba} &=& f_{\ba\bb\bc}\lambda_{1\bb}\lambda_{2\bc}, \nn \\
{\Lambda}_{3\mu\nu\bc} &=& f_{\bc\bm\bn} \big( \lambda_{1\bm}{ \Lambda}_{2\mu\nu\bn} - \lambda_{2\bm}{\Lambda}_{1\mu\nu\bn}\big), \\
{ \Lambda}_{3\mu\ba\bb} &=& 2f_{[\ba|\bm\bn} \big( \lambda_{1\bm}{ \Lambda}_{2\mu\bn|\bb]} - \lambda_{2\bm}{ \Lambda}_{1\mu\bn|\bb]}    \big),\nn\\
{ \Lambda}_{3\ba\bb\bc} &=& 3f_{[\ba|\bm\bn} \big( \lambda_{1\bm}{\Lambda}_{2\bn|\bb\bc]} - \lambda_{2\bm}{\Lambda}_{1\bn|\bb\bc]}    \big).\nn
 \eea
The field strengths that transform homogeneously and with respect to $\lambda_{\ba}$ only, are
\bea\label{B4str}
H_{\mu\nu\rho\sigma\tau} &=& 5 \partial_{[\mu}B_{\nu\rho\sigma\tau]} -10 \kappa^{-1} \cF_{[\mu\nu\bb}B_{\rho\sigma\tau]\bb}, \nn \\
H_{\mu\nu\rho\sigma\ba} &=& 4 \nabla_{[\mu} B_{\nu\rho\sigma]\ba} +6 \kappa^{-1} \cF_{[\mu\nu\bb}B_{\rho\sigma]\bb\ba}, \nn \\
H_{\mu\nu\rho\ba\bb} &=& 3 \nabla_{[\mu} B_{\nu\rho]\ba\bb} - 3\kappa^{-1} \cF_{[\mu\nu \bc}B_{\rho]\bc\ba\bb} +\kappa f_{\ba\bb\bc}B_{\mu\nu\rho\bc}, \\
H_{\mu\nu\ba\bb\bc} &=& 2 \nabla_{[\mu} B_{\nu]\ba\bb\bc} + \kappa^{-1} \cF_{\mu\nu\bd}B_{\bd\ba\bb\bc} -3 \kappa f_{[\ba\bb|\bm}B_{\mu\nu\bm|\bc]}, \nn \\
H_{\mu\ba\bb\bc\bd} &=& \nabla_\mu B_{\ba\bb\bc\bd} + 6 \kappa f_{[\ba\bb|\bm}B_{\mu\bm|\bc\bd]}.\nn
\eea
The Lagrangian, obtained by the dimensional reduction, is a combination of squares of the field strengths \p{B4str}.
In this Section, however, we are interested in the transformations \p{B4trans} and their reducibility properties.

The transformation of the $B_{\mu\nu\rho\sigma}$ with parameter $\Lambda_{\nu \rho\sigma}$ is just the usual transformation of the Abelian $4$-form, which is already known to be of third order of reducibility \cite{KUZ1}. Searching for reducible transformations of non-Abelian tensors, one can note that the consideration about the Yang-Mills field \p{deltaAmured} is also valid and only the transformations with $\lambda_{\ba}=0$ are of interest.

Generalization of a reducible transformation related to the parameter $\Lambda_{\mu\nu\ba}$ can be taken as $\Lambda_{\mu\nu\ba}=2\nabla_{[\mu}\omega_{\nu]\ba}$, so that the first term in $\delta B_{\mu\nu\rho\ba}$ would read $6\nabla_{[\mu}\nabla_{\nu}\omega_{\rho]\ba} = -3 f_{\ba\bb\bc}\cF_{[\mu\nu\bb}\omega_{\rho]\bc}$ and to achieve $\delta B_{\mu\nu\rho\ba}=0$ one should take
\be\label{B4transred1}
\Lambda_{\mu\nu\ba}=2\nabla_{[\mu}\omega_{\nu]\ba}, \;\; \Lambda_{\mu\ba\bb} = \kappa f_{\ba\bb\bc}\omega_{\mu\bc}.
\ee
Parameters $\Lambda_{\mu\nu\ba}$ and $\Lambda_{\mu\ba\bb}$ appear not only in $\delta B_{\mu\nu\rho\ba}$ but also in $\delta B_{\mu\nu\rho\sigma}$ and $\delta B_{\mu\nu\ba\bb}$. In the latter case, contributions from $\Lambda_{\mu\nu\ba}$ and $\Lambda_{\mu\ba\bb}$ cancel each other and reducibility is achieved if one takes $\Lambda_{\ba\bb\bc}=0$. There is no such cancelation for $B_{\mu\nu\rho\sigma}$ transformation, but $\delta B_{\mu\nu\rho\sigma}=0$ if, in addition to \p{B4transred1}, one takes $\Lambda_{\mu\nu\rho} = -3\kappa^{-1} \cF_{[\mu\nu\bb}\omega_{\rho]\bb}$.

In the case of $B_{\mu\nu\ba\bb}$, the straightforward generalization of the Abelian reducible transformation is $\Lambda_{\mu\ba\bb} = \nabla_{\mu}{\omega}_{\ba\bb}$, which generates in $\delta B_{\mu\nu\ba\bb}$ a term
\be\label{B4transred2}
-2 f_{[\ba |\bm\bn} \cF_{\mu\nu\bm}{\omega}_{\bn|\bb]} \neq 0.
\ee
Though it looks structurally different from other terms in $\delta B_{\mu\nu\ba\bb}$, it can be rewritten using the identity
\be\label{B4transred3}
-2f_{[\ba|\bm\bn}{\omega}_{\bn|\bb]} =3 f_{[\bm \ba|\bn}{\omega}_{\bn|\bb]}+f_{\ba\bb\bn}{\omega}_{\bm\bn}.
\ee
Therefore, to obtain $\delta B_{\mu\nu\ba\bb}=0$ one has to choose
\be\label{B4transred4}
\Lambda_{\mu\ba\bb} = \nabla_{\mu}{\omega}_{\ba\bb}, \;\; \Lambda_{\mu\nu\ba} = \kappa^{-1} \cF_{\mu\nu\bb}{\omega}_{\bb\ba}, \;\; \Lambda_{\ba\bb\bc} = -3\kappa f_{[\ba\bb|\bm}{\omega}_{\bm|\bc]}.
\ee
Other variations \p{B4trans} vanish after this substitution.

Parameters $\omega_{\mu\nu\ba}$ and ${\omega}_{\nu\ba\bb}$ induce two independent trivial transformations. Their linear combination together with trivial $\Lambda_{\mu\nu\rho}$ transformation
\bea\label{B4transred5}
\Lambda^{triv}_{\mu\nu\rho} &=& 3 \partial_{[\mu} {\omega}_{\nu\rho]} -3\kappa^{-1} \cF_{[\mu\nu\bb}\omega_{\rho]\bb}, \nn \\
\Lambda^{triv}_{\mu\nu\ba} &=& 2 \nabla_{[\mu} \omega_{\nu]\ba} - \kappa^{-1}\cF_{\mu\nu\bb}{\omega}_{\ba\bb}, \nn \\
\Lambda^{triv}_{\mu\ba\bb}  &=& \nabla_\mu {\omega}_{\ba\bb} + \kappa f_{\ba\bb\bc} \omega_{\mu\bc}, \\
\Lambda^{triv}_{\ba\bb\bc} &=& -3 \kappa f_{[\ba\bb |\bd} {\omega}_{\bd|\bc]}\nn
\eea
is the general trivial transformation.

As a result, one can say that the parameters $\Lambda$ are defined to up the gauge transformations $\Lambda^\prime \sim \Lambda + \delta\Lambda$, where
\bea\label{B4deltaLambda}
\delta\Lambda_{\mu\nu\rho} &=& 3 \partial_{[\mu} { \omega}_{\nu\rho]} -3\kappa^{-1} \cF_{[\mu\nu\bb}\omega_{\rho]\bb}, \nn \\
\delta\Lambda_{\mu\nu\ba} &=& 2 \nabla_{[\mu} \omega_{\nu]\ba} - \kappa^{-1}\cF_{\mu\nu\bb}{\omega}_{\ba\bb}, \nn \\
\delta\Lambda_{\mu\ba\bb}  &=& \nabla_\mu {\omega}_{\ba\bb} + \kappa f_{\ba\bb\bc} \omega_{\mu\bc}, \\
\delta\Lambda_{\ba\bb\bc} &=& -3 \kappa f_{[\ba\bb |\bd} {\omega}_{\bd|\bc]}.\nn
\eea
Remarkably, \p{B4deltaLambda} resembles, up to change of notations, transformations of $B$ fields in the $3$-form case. Therefore, the results of the previous Section imply that \p{B4deltaLambda} are themselves reducible, and gauge transformations of $\omega$ are
\bea\label{B4deltaomega}
\delta \omega_{\mu\nu} &=& \kappa^{-1} \cF_{\mu\nu\ba}\sigma_\ba +
\partial_{\mu}\zeta_{\nu} - \partial_\nu \zeta_\mu, \nn \\
\delta\omega_{\mu\ba} &=& \nabla_{\mu}\sigma_{\ba}, \\
\delta\omega_{\ba\bb} &=& -\kappa f_{\ba\bb\bc}\sigma_\bc.\nn
\eea
The degree of reducibility of \p{B4transred5} is, therefore, one more than \p{Btrans}, that is $2$ for non-Abelian and $3$ for Abelian transformations, as the $\zeta_\mu$ transformations are reducible again, $\delta\zeta_\mu = \partial_\mu \zeta$.

It should be noted that, just as in the $3$-form case,
\p{B4transred5} is the most general trivial transformation, though
the analysis is more subtle than in the case of the $3$-form.  One
can try to find more general transformation by solving condition
$\delta B_{\ba\bb\bc\bd}=0$ \p{B4trans} with $\lambda_{\bm}=0$.
Multiplying this condition by a structure constant of the algebra
and using the Jacobi identity, one can obtain a relation
 \be\label{B4transred6}
 2\Lambda_{\ba\bb\bc} +
3f_{\bp[\ba\bb}\Lambda_{\bc]\bm\bn}f_{\bp\bm\bn} =0.
 \ee
Therefore, for trivial transformations $\Lambda_{\ba\bb\bc} =
f_{\bd[\ba\bb}{\check\omega}_{\bc]\bd}$. Substituting this relation
back into $\delta B_{\ba\bb\bc\bd}=0$ \p{B4trans}, one can find that
the antisymmetric part of ${\check\omega}_{\bc\bd}$, which
corresponds to the already known transformation \p{B4transred5}, is
not constrained, while the symmetric part is subjected to the
relation
 \be\label{B4checkomega}
 \big(f_{\ba\bb\bm}f_{\bc\bd\bn}-f_{\ba\bc\bm}f_{\bb\bd\bn}+f_{\ba\bd\bm}f_{\bb\bc\bn}
\big)\big( {\check\omega}_{\bm\bn} + {\check\omega}_{\bn\bm}
\big)=0.
 \ee
 Though \p{B4checkomega} is nontrivial and is solved by
${\check\omega}_{\bm\bn} + {\check\omega}_{\bn\bm} \sim
\delta_{\bm\bn} {\check\omega}$, this solution is not compatible
with \p{B4trans}. It is worth noting that when quantizing the entire
theory, the scalar field does not cause problems and such a
symmetry can be just gauge-fixed by removing some of the scalar
fields.

Just like the $3$-form case, some of the non-Abelian fields acquire mass terms and others become St\"uckelberg fields, the latter restoring the gauge symmetries of the former. The structure of mass terms and St\"uckelberg fields for $B_{\mu\nu\rho\ba}$ and $B_{\mu\nu\ba\bb}$ tensors is similar to $B_{\mu\nu\ba}$ and $B_{\mu\ba\bb}$ in the $3$-form case, with $S_{\mu\nu\ba}=f_{\ba\bb\bc}B_{\mu\nu\bb\bc}$  and $S_{\mu\ba\bb}=B_{\mu[\ba|\bm\bn}f_{\bm\bn|\bb]}$ serving as compensators for $B_{\mu\nu\rho\ba}$ and $B_{\mu\nu\ba\bb}$ gauge symmetries, while for scalar fields it is more complicated.

The similarity between transformations in the $3$-form system and
the reduced transformations \p{B4transred5} closely mimics the
properties of the corresponding Abelian $3$-form, and one can expect
that this analogy extends to the general $p$-form case.

\section{General antisymmetric tensor field in arbitrary dimension}
Considerations that were used in the reduction of the $3$- and $4$-form fields
can be straightforwardly extended to the case of antisymmetric tensor field of
arbitrary rank. They can be further simplified if one notes that
for the purpose of study of transformation laws one does not even
need the action\footnote{Such an action is derived the same way as
it was done for antisymmetric fields of third and fourth ranks.}.
Moreover, the general structure of field transformations and
redefinitions can be understood on the base of  the 3-form case up
to coefficients.

The Abelian antisymmetric tensor of the rank $r+1$ in a
higher-dimensional curved spacetime transforms as
 \bea\label{genCtrans}
\delta C_{M_1 M_2 \ldots M_{r+1}} &=&\xi{}^Q \partial_Q C_{M_1 M_2 \ldots M_{r+1}} + \partial_{M_1} \xi^Q \, C_{Q M_2 \ldots M_{r+1}} + \partial_{M_2}\xi^Q \,C_{M_1 Q M_3 \ldots M_{r+1}}+ \ldots \nn \\
 &&+\ldots + \partial_{M_{r+1}}\xi^Q \, C_{M_1 M_2 \ldots M_{r}Q}+ (r+1) \partial_{[M_1}\Lambda_{M_2 \ldots M_{r+1}]}.
 \eea
To perform the dimensional reduction, we assume the metric on the curved
space takes form \p{Gstruct} and apply the prescription from the
Section $2$ to the form $C_{M_1 M_2 \ldots M_{r+1}}$ and the
parameters $\xi{}^Q$ and $\Lambda_{M_1 \ldots M_{r}}$, so that the
tensor $C_{M_1 M_2 \ldots M_{r+1}}$ decomposes into
 \bea\label{genCexp}
C_{\mu_1 \mu_2 \ldots \mu_{r+1}}(x), \;\; C_{\mu_1 \mu_2 \ldots \mu_{r}\ba_1}(x)\, e_{a_1 \ba_1}, \;\; \ldots C_{\mu_1 \ldots \mu_n \ba_1 \ldots \ba_{r-n+1}}(x)\, e_{a_1 \ba_1}\ldots e_{a_{r-n+1}\ba_{r-n+1}}, \;\; \ldots, \nn \\
C_{\mu_1 \ba_1 \ldots \ba_{r}}(x)\, e_{a_1 \ba_1}\ldots e_{a_{r}\ba_{r}}, \;\; C_{ \ba_1 \ldots \ba_{r+1}}(x)\, e_{a_1 \ba_1}\ldots e_{a_{r+1}\ba_{r+1}},
 \eea
though transformation laws of $C$ fields contain derivatives of
$\lambda_\ba$. The fields $ B_{\mu_1 \ldots \mu_n \ba_1 \ldots
\ba_{r-n+1}}(x)$ which transform covariantly with respect to
$\lambda_\ba$ can be found using the prescription
 \be\label{genBC}
B_{{\bar M}_1 {\bar M}_2 \ldots {\bar M}_{r+1}} = \big( E^{-1} \big)_{{\bar M}_1}{}^{M_1 }\big( E^{-1} \big)_{{\bar M}_2}{}^{M_2 } \ldots \big( E^{-1} \big)_{{\bar M}_{r+1}}{}^{M_{r+1} }C_{M_1 M_2 \ldots M_{r+1}}.
 \ee
It can be noted that, as the transformation law of $C_{\mu_1 \ldots \mu_n \ba_1 \ldots \ba_{r-n+1}}(x)$ contains at most one derivative, the transformation of $B_{\mu_1 \ldots \mu_n \ba_1 \ldots \ba_{r-n+1}}(x)$ can only involve a term with covariant derivative of some parameter, product of the Yang-Mills field strength with another parameter and a third parameter without any extra fields and derivatives. Calculation of $\delta B_{\mu_1 \ldots \mu_n \ba_1 \ldots \ba_{r-n+1}}(x)$ gives
\bea\label{genBtrans}
\delta B_{\mu_1 \ldots \mu_n \ba_1 \ldots \ba_{r-n+1}}&=& (r-n+1) f_{[\ba_1 |\bb\bc} \lambda_{\bb} B_{\mu_1 \ldots \mu_n \bc| \ba_2 \ldots \ba_{r-n+1}]} + n \nabla_{[\mu_1} \Lambda_{\mu_2 \ldots \mu_n] \ba_1 \ldots \ba_{r-n+1}} + \nn \\
&&+\frac{1}{2}\kappa^{-1} n (n-1)(-1)^n \cF_{[\mu_1\mu_2\bb}\Lambda_{\mu_3\ldots \mu_n]\bb \ba_1 \ldots \ba_{r-n+1}}- \\
&&- \frac{1}{2}\kappa (-1)^n (r-n+1)(r-n)f_{[\ba_1 \ba_2 |\bc}\Lambda_{\mu_1 \ldots \mu_n \bc | \ba_3 \ldots \ba_{r-n+1}]}.\nn
\eea
As can be expected, the transformations \p{genBtrans} with parameters $\big\{ \lambda_{\ba} , \Lambda_{\mu_1 \ldots \mu_r \ba}, \ldots \Lambda_{\ba_1 \ldots \ba_{r+1}}\big\}$ are non-Abelian, with commutator of $\delta_2$ and $\delta_1$ given by a transformation with parameters
\bea\label{genBtranscom}
\lambda_{3\ba} &=& f_{\ba\bb\bc}\lambda_{1\bb}\lambda_{2\bc}, \nn \\
{\Lambda}_{3\mu_1 \ldots  \mu_r \ba} &=& f_{\ba\bm\bn} \big( \lambda_{1\bm}{ \Lambda}_{2\mu_1 \ldots  \mu_r\bn} - \lambda_{2\bm}{\Lambda}_{1\mu_1 \ldots  \mu_r\bn}\big),  \ldots \\
{ \Lambda}_{3\mu_1 \ldots \mu_n \ba_1 \ldots \ba_{r-n+1}} &=& (r-n+1)f_{[\ba|\bm\bn} \big( \lambda_{1\bm}{ \Lambda}_{2\mu_1 \ldots \mu_n  \bn|\ba_2 \ldots  \ba_{r-n+1}]} - \lambda_{2\bm}{ \Lambda}_{1\mu_1 \ldots \mu_n \bn|\ba_2 \ldots  \ba_{r-n+1}]}    \big), \ldots,\nn\\
{ \Lambda}_{3\ba_1 \ldots \ba_{r+1}} &=& (r+1)f_{[\ba|\bm\bn} \big( \lambda_{1\bm}{ \Lambda}_{2\bn|\ba_2 \ldots \ba_{r+1}]} - \lambda_{2\bm}{\Lambda}_{1\bn|\ba_2 \ldots \ba_{r+1}]}    \big).\nn
\eea

The $\Lambda$ transformations \p{genBtrans} leave invariant the field strength
\bea\label{genBstr}
H_{\mu_1 \ldots \mu_{n+1} \ba_1 \ldots \ba_{r-n+1}}&=&  (n+1 ) \nabla_{[\mu_1} B_{\mu_2 \ldots \mu_{n+1} ]\ba_1 \ldots \ba_{r-n+1}}-\nn \\
&&- \frac{1}{2}\kappa^{-1}(-1)^n n (n+1) \cF_{[\mu_1\mu_2\bb}B_{\mu_3\ldots \mu_{n+1}]\bb \ba_1 \ldots \ba_{r-n+1}}+\nn \\
&&+ \frac{1}{2}\kappa (-1)^n (r-n+1)(r-n)f_{[\ba_1 \ba_2 |\bc}B_{\mu_1 \ldots \mu_{n+1} \bc | \ba_3 \ldots \ba_{r-n+1}]}.
\eea

Reducibility of transformations \p{genBtrans} can be shown in the following way. As was discussed in the Section 3, only the transformations with $\lambda_\ba=0$ can be reducible. Let us take the transformation \p{genBtrans} of the tensor of particular rank $n$, $B_{\mu_1 \ldots \mu_n \ba_1 \ldots \ba_{r-n+1}}$, which involves the tensorial parameters of ranks $n-1$, $n-2$ and $n$, and find what nontrivial parameters reduce it to zero, generalizing trivial Abelian transformation with parameter of rank $n-1$
\be\label{genLambdaomegan1}
\Lambda_{\mu_2 \ldots \mu_{n}\ba_1 \ldots \ba_{r-n+1}} = (n-1)\nabla_{[\mu_2} \omega_{\mu_3 \ldots \mu_{n}]\ba_1 \ldots \ba_{r-n+1}}.
\ee
This, despite the presence of $n$, should not be considered at this point as a general expression valid for all $n$. In particular, it should be substituted into the first term of \p{genBtrans} only.

After substitution, one can evaluate commutator of covariant derivatives and, noting that
\bea\label{genId}
&&\frac{1}{2} (r-n+2)(r-n+1)f_{[ \bb \ba_1 | \bc} \omega_{\mu_3 \ldots \mu_n \bc | \ba_2 \ldots \ba_{r-n+1}]} = \nn \\
&&=-\frac{1}{2} (r-n)(r-n+1) f_{[\ba_1 \ba_2 |\bc} \omega_{\mu_3 \ldots \mu_n \bb \bc | \ba_3 \ldots \ba_{r-n+1}]} - \\
&&-(r-n+1) f_{[\ba_1 |\bb\bc} \omega_{\mu_3 \ldots \mu_n \bc | \ba_2 \ldots \ba_{r-n+1}]},\nn
\eea
one may find that $\delta B_{\mu_1 \ldots \mu_n \ba_1 \ldots \ba_{r-n+1}}=0$ if other parameters read
\bea\label{genLambdaomegan2}
\Lambda_{\mu_3 \ldots \mu_n \bb \ba_1 \ldots \ba_{r-n+1}} = - \frac{1}{2}\kappa (-1)^{n-2} (r-n+2)(r-n+1) f_{[\bb\ba_1 |\bc}\omega_{\mu_3\ldots \mu_n \bc| \ba_2 \ldots \ba_{r-n+1}]},   \\
\Lambda_{\mu_1 \ldots \mu_n \bc \ba_3 \ldots \ba_{r-n+1}} = +\frac{1}{2}\kappa^{-1}(-1)^n n(n-1) \cF_{[\mu_1 \mu_2 \bb}\omega_{\mu_3 \ldots \mu_n] \bb\bc \ba_3 \ldots \ba_{r-n+1}}.          \label{genLambdaomegan}
\eea
The parameters \p{genLambdaomegan1}, \p{genLambdaomegan2}, \p{genLambdaomegan} enter the transformation laws of other tensors with ranks $n-1$, $n+1$, $n-2$, $n+2$. Substituting \p{genLambdaomegan2} and \p{genLambdaomegan1} into relation $\delta B_{\mu_1 \mu_{n-1} \ba_1 \ldots \ba_{r-n+2}}=0$, one can find that they mutually cancel each other and one can take the remaining parameter of rank $n-3$ to zero.
The transformation law of the tensor of rank $n-2$ contains parameters with ranks $n-3$, $n-4$ and $n-2$. As the first is zero and the latter after substitution \p{genLambdaomegan2} results in a trivial contribution, the remaining parameter of rank $n-4$ can be taken zero. Therefore, transformations \p{genLambdaomegan1}, \p{genLambdaomegan2}, \p{genLambdaomegan} do not ``propagate'' to the bottom of the tower. The same happens to $n+1$ and $n+2$ rank tensors, so \p{genLambdaomegan1}, \p{genLambdaomegan2}, \p{genLambdaomegan} provide a self-contained trivial transformation, ensuring that \p{genBtrans} are at least the first order reducible.

Obviously, this consideration applies to the tensors of other ranks, and the general trivial transformation should be a linear combination of \p{genLambdaomegan1}, \p{genLambdaomegan2}, \p{genLambdaomegan} for different $n$. In particular, transformations of $B_{\mu_1 \ldots \mu_{n-1} \ba_1 \ldots \ba_{r-n+2}}$, $B_{\mu_1 \ldots \mu_{n-2} \ba_1 \ldots \ba_{r-n+3}}$, $B_{\mu_1 \ldots \mu_{n+1} \ba_1 \ldots \ba_{r-n}}$, $B_{\mu_1 \ldots \mu_{n+2} \ba_1 \ldots \ba_{r-n-1}}$ also induce transformation of $B_{\mu_1 \ldots \mu_n \ba_1 \ldots \ba_{r-n+1}}$. Combined reducible transformation of $B_{\mu_1 \ldots \mu_n \ba_1 \ldots \ba_{r-n+1}}$ is given by the parameters
\bea\label{genLambdaomegaall}
\Lambda^{triv}_{\mu_1 \ldots \mu_{n-1} \ba_1 \ldots \ba_{r-n+1}} &=& (n-1) \nabla_{[\mu_1} \omega^{(n)}_{\mu_2 \ldots \mu_{n-1}]\ba_1 \ldots \ba_{r-n+1}} +  \nn \\
&&+\frac{1}{2}\kappa^{-1}(n-1)(n-2)(-1)^{n-1} \cF_{[\mu_1 \mu_2 \bb}\omega^{(n-1)}_{\mu_3 \ldots \mu_{n-1}]\bb \ba_1 \ldots \ba_{r-n+1}}- \nn \\
&&-\frac{1}{2}\kappa (-1)^{n+1} (r-n+1)(r-n) f_{[\ba_1 \ba_2 |\bc} \omega^{(n+1)}_{\mu_1 \ldots \mu_{n-1} \bc | \ba_3 \ldots \ba_{r-n+1}]}, \nn \\
\Lambda^{triv}_{\mu_1 \ldots \mu_{n-2} \ba_1 \ldots \ba_{r-n+1}} &=& (n-2) \nabla_{[\mu_1} \omega^{(n-1)}_{\mu_2 \ldots \mu_{n-2}] \ba_1 \ldots \ba_{r-n+2}} + \nn \\
&& + \frac{1}{2} \kappa^{-1} (n-2)(n-3) (-1)^{n-2}  \cF_{[\mu_1 \mu_2 \bb}\omega^{(n-2)}_{\mu_3 \ldots \mu_{n-2}]\bb \ba_1 \ldots \ba_{r-n+2}} +  \\
&& - \frac{1}{2}\kappa (-1)^{n-2} (r-n+2)(r-n+1) f_{[\ba_1 \ba_2 |\bc} \omega^{(n)}_{\mu_1 \ldots \mu_{n-2} \bc | \ba_3 \ldots \ba_{r-n+1}]}, \nn \\
\Lambda^{triv}_{\mu_1 \ldots \mu_{n} \ba_1 \ldots \ba_{r-n+1}} &=& n \nabla_{[\mu_1} \omega^{(n+1)}_{\mu_2 \ldots \mu_{n}] \ba_1 \ldots \ba_{r-n}} + \nn \\
&& + \frac{1}{2} \kappa^{-1} n(n-1) (-1)^{n}  \cF_{[\mu_1 \mu_2 \bb}\omega^{(n)}_{\mu_3 \ldots \mu_{n}]\bb \ba_1 \ldots \ba_{r-n}} + \nn \\
&& - \frac{1}{2}\kappa (-1)^{n} (r-n)(r-n-1) f_{[\ba_1 \ba_2 |\bc} \omega^{(n+2)}_{\mu_1 \ldots \mu_{n} \bc | \ba_3 \ldots \ba_{r-n}]}. \nn
\eea
Here $\omega^{(k)}$ are the parameters of trivial $B_{\mu_1 \ldots \mu_k \ba_1 \ldots \ba_{r-k+1}}$ transformation \p{genLambdaomegan1}, \p{genLambdaomegan2}, \p{genLambdaomegan}.

It is important to note that $\Lambda^{triv}_{\mu_1 \ldots \mu_{n-1} \ba_1 \ldots \ba_{r-n+1}}$ coincides with $\delta B_{\mu_1 \ldots \mu_{n-1} \ba_1 \ldots \ba_{r-n+1}}$ \p{genBtrans} with parameters $\omega^{(n)}_{\mu_2 \ldots \mu_{n-1}\ba_1 \ldots \ba_{r-n+1}}$, $\omega^{(n-1)}_{\mu_3 \ldots \mu_{n-1}\bb \ba_1 \ldots \ba_{r-n+1}}$, $ \omega^{(n+1)}_{\mu_1 \ldots \mu_{n-1} \bc | \ba_3 \ldots \ba_{r-n+1}}$. Similarly, $\Lambda^{triv}_{\mu_1 \ldots \mu_{n-2} \ba_1 \ldots \ba_{r-n+1}}$ and $\Lambda^{triv}_{\mu_1 \ldots \mu_{n} \ba_1 \ldots \ba_{r-n+1}}$ coincide with $\delta B_{\mu_1 \ldots \mu_{n-2} \ba_1 \ldots \ba_{r-n+2}}$ and $\delta B_{\mu_1 \ldots \mu_{n} \ba_1 \ldots \ba_{r-n}}$. As \p{genBtrans} are known to be reducible, with parameters of trivial transformation given by \p{genLambdaomegaall}, one can find nontrivial $\omega$ in \p{genLambdaomegaall} so $\Lambda^{triv}_{\mu_1 \ldots \mu_{n-1} \ba_1 \ldots \ba_{r-n+1}} =0$, $\Lambda^{triv}_{\mu_1 \ldots \mu_{n-2} \ba_1 \ldots \ba_{r-n+1}}=0$, $\Lambda^{triv}_{\mu_1 \ldots \mu_{n} \ba_1 \ldots \ba_{r-n+1}}=0$ and it is possible to reduce \p{genLambdaomegaall} once again.

It should be noted that another non-Abelian system of antisymmetric
tensor fields exists which includes one dynamical $n$-form, one
St\"uckelberg $n-1$-form and the Yang-Mills field \cite{nishino}.
Their transformation laws are given by
 \bea\label{nishgen21}
\delta B_{\mu_1 \ldots \mu_n \ba} = n \nabla_{[\mu_1} \beta_{\mu_2 \ldots \mu_n]\ba} -\frac{1}{2}n(n-1) f_{\ba\bb\bc}\cF_{[\mu_1\mu_2\bb}\gamma_{\mu_3\ldots \mu_n]\bc} + f_{\ba\bb\bc}\lambda_\bb B_{\mu_1 \ldots \mu_n \bc}, \nn \\
\delta C_{\mu_1 \ldots \mu_{n-1}\ba} = \beta_{\mu_1\ldots \mu_n \ba} + (n-1)\nabla_{[\mu_1} \gamma_{\mu_2 \ldots \mu_{n-1}] \ba}, \;\;
\delta A_{\mu\ba} = \partial_\mu \lambda_\ba + f_{\ba\bb\bc}\lambda_\bb A_{\mu\bc}.
 \eea
These transformations are reducible. Assuming that $\lambda_\bb=0$ and
 \be\label{nishgen22}
\beta_{\mu_2 \ldots \mu_n \ba} = (n-1) \nabla_{[\mu_2} \omega_{\mu_3 \ldots \mu_n] \ba},
 \ee
which is a direct generalization of an Abelian relation, one finds that if
 \be\label{nishgen23}
\gamma_{\mu_3 \ldots \mu_n\ba} = -\omega_{\mu_3 \ldots \mu_n\ba}
 \ee
transformations of $B_{\mu_1 \ldots \mu_n \ba}$ and $C_{\mu_1 \ldots \mu_{n-1}\ba}$ vanish.

While the Abelian $p$-form theories are $p-1$ times reducible, in
the special case noted in \cite{nishino}, situation is different:
due to \p{nishgen23}, there is no nontrivial $\omega_{\mu_1 \ldots
\mu_{n-2}\ba}$ for which $\gamma_{\mu_1 \ldots \mu_{n-2}\ba}$
vanishes. Therefore, unlike the Abelian case, the non-Abelian model
of $n$-forms \cite{nishino} is only the first stage reducible.

\section{Summary}
In this paper we have analyzed the field models obtained by the
dimensional reduction of the free antisymmetric tensor fields
minimally coupled to gravity, with the group as a compact
submanifold, and have discussed the gauge structure of the
resulting theories. We have considered the ranks $3$, $4$
tensor models and then studied the arbitrary rank tensor
field. All these models are the reducible gauge theories
possessing rich spectrum of consistently interacting fields
including non-Abelian totally antisymmetric tensors, vectors and
scalars.  The dimensionful parameter related to the size of the
compact submanifold plays the role of the coupling constant.

The main focus of our study was the properties of gauge
transformations of the resulting theories. It has been shown
that the set of all gauge transformations as a whole forms a
non-Abelian algebra, including the Abelian subalgebra  of
transformations for higher-rank tensor fields. It is important to
emphasize that the reduction procedure leads to gauge-invariant
formulations of the Lagrangian, where higher-rank tensor fields
contain mass terms, which generally indicates the presence of
St\"uckelberg fields in the composition of the fields. Indeed, in all the considered cases, the transformation laws of the
non-Abelian fields included combinations of parameters without
derivatives, allowing one to define the St\"uckelberg fields, which
transformation laws are just shifts with respect to some of the
parameters. These St\"uckelberg fields ensure that the theory is
gauge-symmetric in spite of presence of the mass terms for other
non-Abelian tensor fields and, in principle, can be gauged away. It
is the presence of the mass terms in the system under consideration
that allows to efficiently bypass the no-go theorems
\cite{henneaux1}, \cite{bekaert1}, where the mass deformations of
the free theory were not considered.

We found that all the considered gauge transformations are
reducible and that the reducibility stages coincide with ones of
corresponding free Abelian theories, with additional reducibility in
the vector and scalar sectors of the models. Moreover, the
non-Abelian theories of antisymmetric tensors obtained by the
dimensional reduction obligatory contain the St\"uckelberg fields.

The obtained results open new directions in the study of the non-Abelian theories involving $p$-forms.
\begin{itemize}
\item{Though only the simplest $p$-form models minimally coupled to
gravity were discussed in this paper, the method is
certainly not limited to these cases and can be applied to
interacting theories and self-dual actions of Pasti-Sorokin-Tonin
type \cite{PST} in higher dimensions. Results of the paper also
should be valid in these cases, as the actions themselves played
little role in the construction and the gauge transformation laws of
the fields can be obtained from higher-dimensional analogs through
the reduction prescriptions.}
\item{It would be of interest to understand whether the results of this
paper extend to the systems obtained by the dimensional reduction on
manifolds other than groups, such sphere reductions of $d=10$ and
$d=11$ supergravities.}
\item{As is well known, the free $p$-form fields in various dimensions are
related to one another by the duality transformations. It is natural
to expect that some kind of dualities should take place for
interacting non-Abelian $p$-form theories.}
\item{It is of interest to construct supersymmetric versions of some of the considered models.}
\end{itemize}

Also the obtained results should be useful in studying of the quantum
structure of the non-Abelian tensor theories:
\begin{itemize}
\item{Canonical quantization. Construction of canonical formulation, derivation of the constraints in the
phase space and their algebra, carrying out BRST-BFV quantization
\cite{BFV1}, \cite{BFV2} (see also book \cite{HT}).}
\item{Covariant quantization. Derivation of the path integral with correct set of gauge conditions
and ghost fields using BV approach \cite{BV} (see also book
\cite{HT}).}
\item{Studying the divergence structure and deriving quantum effective actions.}
\end{itemize}
We plan to explore these issues in forthcoming works.

\end{document}